 \documentstyle[12pt]{article}

\let\b=\beta \let\a=\alpha
\let\l=\lambda

  \let\la=\label  
  
\def\nn{\nonumber} \def\bd{\begin{document}} \def\ed{\end{document}}
\def\ds{\documentstyle} \let\fr=\frac \let\bl=\bigl \let\br=\bigr
\let\Br=\Bigr \let\Bl=\Bigl \let\bm=\bibitem \let\na=\nabla
\let\pa=\partial \let\ov=\overline \newcommand{\be}{\begin{equation}}
\newcommand{\ee}{\end{equation}} \def\ba{\begin{array}} \def\ea{\end{array}}
\newcommand{\ho}[1]{$\, ^{#1}$} \newcommand{\hoch}[1]{$\, ^{#1}$}
\newcommand{\bea}{\begin{eqnarray}} \newcommand{\eea}{\end{eqnarray}}
\newcommand{\ra}{\rightarrow} \newcommand{\lra}{\longrightarrow}
\newcommand{\Lra}{\Leftrightarrow} \newcommand{\ap}{\alpha^\prime}
\newcommand{\bp}{\beta^\prime} \newcommand{\tr}{{\rm tr} }
\newcommand{\Tr}{{\rm Tr} } \newcommand{\NP}{Nucl. Phys. }
\newcommand{\tamphys}{\it Center for Theoretical Physics\\
Physics Department \\ Texas A \& M University
\\ College Station, Texas 77843}

\newcommand{\auth}{ M. J. Duff\hoch{\dagger} and J. Rahmfeld}

\begin{document}
\baselineskip=20pt
\hsize=6.5truein
\hoffset=-0.75truein
\hfill{}

\hfill{CTP-TAMU-25/94}

\hfill{hep-th/9406105}

\vspace{24pt}

\begin{center}
{\large {\bf MASSIVE STRING STATES AS EXTREME BLACK HOLES}}

\vspace{36pt}

\auth

\vspace{10pt}

{\tamphys}

\vspace{44pt}

\underline{ABSTRACT}

\end{center}

We consider the Schwarz-Sen spectrum of elementary electrically charged
massive $N_R=1/2$ states of the four-dimensional heterotic string and show
the maximum spin $1$ supermultiplets to correspond to extreme black hole
solutions.  The $N_L=1$ states and $N_L>1$ states (with vanishing left-moving
internal momentum) admit a single scalar-Maxwell description with
parameters $a=\sqrt 3$ or $a=1$, respectively. The corresponding
solitonic magnetically charged spectrum conjectured by Schwarz and Sen on
the basis of $S$-duality is also described by extreme black
holes.

{\vfill
\leftline{JUNE 94, revised DECEMBER 94}
\vskip 10pt
\footnoterule
{\footnotesize
 \hoch{\dagger} Research
 supported in part by NSF Grant PHY-9106593.}}
\baselineskip=20pt

\pagebreak

\setcounter{page}{1}

The idea that elementary particles might behave like black holes is not a
new one \cite{Salam}. Intuitively, one might expect that
a pointlike object whose mass exceeds the Planck mass, and whose Compton
wavelength is therefore less than its Schwarzschild radius, would exhibit
an event horizon. In the absence of a consistent quantum theory of gravity,
however, such notions would always remain rather vague. Superstring theory,
on the other hand, not only predicts such massive states but may provide us
with a consistent framework in which to discuss them. The purpose of the
present paper is to confirm the claim \cite {Khuri} that
certain massive excitations of four-dimensional superstrings are indeed
black holes.  Of course, non-extreme black holes would be unstable due to
the Hawking effect. To describe stable elementary particles, therefore, we
must focus on extreme black holes whose masses saturate a Bogomol'nyi bound
\footnote{ The relationship between extremal black holes and the
gravitational field around some of the elementary string states has also been
discussed in \cite{Fabbrichesi} and \cite{Sen2}.}.  The present paper
therefore remains agnostic concerning the stronger claims\cite{Ellis,Russo}
that {\it all} black holes are single string states or, conversely, that all
massive string states are black holes.

Specifically, we shall consider the four-dimensional heterotic string
obtained by toroidal compactification. At a generic point in the moduli
space of vacuum configurations the unbroken gauge symmetry is $U(1)^{28}$
and the low energy effective field theory is described by $N=4$ supergravity
coupled to $22$ abelian vector multiplets. A recent paper \cite{Khuri}
showed that this theory exhibits both electrically and magnetically charged
black hole solutions corresponding to scalar-Maxwell parameter
$a=0,1,\sqrt3$. In other words, by choosing appropriate combinations of
dilaton and moduli fields to be the scalar field $\phi$ and appropriate
combinations of the field strengths and their duals to be the Maxwell field
$F$, the field equations can be consistently truncated to a form given by
the Lagrangian
\be
{\cal L}= \frac{1}{32\pi}\sqrt{-g}\left [R-\frac{1}{2}(\partial
\phi)^2-\frac{1}{4}e^{-a \phi}F^2 \right]
\la{1}
\ee
for these three values of $a$. (A {\it consistent} truncation is defined to
be one for which all solutions of the truncated theory are solutions of the
original theory). In the case of zero angular momentum, the bound  between
the black hole ADM mass $m$, and the electric charge  $Q=\int
e^{-a\phi}{}{\tilde F}/8\pi$, where a tilda denotes the dual, is given by
\be
m^2\geq Q^2/4(1+a^2)
\la{2}
\ee
where, for simplicity, we have set the asymptotic value of $\phi$ to zero.
The $a=0$ case yields the Reissner-Nordstrom solution which, notwithstanding
contrary claims in the literature, does solve the low-energy string
equations. The $a=1$ case yields the dilaton black hole
\cite{Gibbons,Garfinkle}. The $a=\sqrt{3}$ case corresponds to the
Kaluza-Klein black hole and the "winding" black hole \cite{Khuri} which are
related to each other by $T$-duality. The Kaluza-Klein solution has been
known for some time \cite{Gibbons} but only recently recognized \cite{Khuri}
as a heterotic string solution.

Let us denote by $N_L$ and $N_R$ the number of left and right oscillators
respectively. We shall consider the Schwarz-Sen \cite{Schwarz} $O(6,22;Z)$
invariant spectrum of elementary electrically charged massive $N_R=1/2$
states of this four-dimensional heterotic string, and show that the spin
zero states correspond to extreme limits of  black hole solutions which
preserve [A$1/2$ of the spacetime supersymmetries. By supersymmetry, the
black hole interpretation then applies to all members of the $N=4$
supermultiplet \cite{Gibbons2,Aichelburg}, which has $s_{max}=1$. For a
subset of states the low-energy string action can be truncated to (\ref{1}).
 The scalar-Maxwell parameter is given by $a=\sqrt 3$ for $N_L=1$ and $a=1$
for $N_L>1$ (and vanishing left-moving internal momenta). The other states
with $N_L>1$  are extreme black holes too, but are not described by a single
scalar truncation of the type (\ref{1}). The $N=4$ supersymmetry algebra
possesses two central charges $Z_1$ and $Z_2$.  The $N_R=1/2$ states
correspond to that subset of the full spectrum that belong to the $16$
complex dimensional ($s_{max}\geq1$) representation of the $N=4$
supersymmetry algebra, are annihilated by half of the supersymmetry
generators and saturate the strong Bogomol'nyi bound $m=|Z_1|=|Z_2|$. As
discussed in \cite{Witten,Schwarz}, the reasons for focussing on this N=4
theory, aside from its simplicity, is that one expects that the allowed
spectrum of electric and magnetic charges is not renormalized by quantum
corrections, and that the allowed mass spectrum of particles saturating the
Bogomol'nyi bound is not renormalized either.


Following \cite{Font} (see also \cite{Duff}), Schwarz and Sen have also
conjectured \cite{Schwarz} on the basis of string/fivebrane duality
\cite{Duff1} that, when the solitonic excitations are included, the full
string spectrum is invariant not only under the target space $O(6,22;Z)$
($T$-duality) but also under the strong/weak coupling $SL(2,Z)$
($S$-duality). The importance of $S$-duality in the context of black holes
in string theory has also been stressed in \cite{Kalara}.  Schwarz and Sen
have constructed a manifestly $S$ and $T$ duality invariant mass spectrum.
$T$-duality transforms electrically charged winding states into electrically
charged Kaluza-Klein states, but $S$-duality transforms elementary
electrically charged string states into solitonic monopole and dyon states.
We shall show that these states are also described by the extreme
magnetically charged black hole solutions.
 Indeed, although the results of the present paper may be
understood without resorting to string/fivebrane duality, it nevertheless
provided the motivation. After compactification from $D=10$ dimensions to
$D=4$, the solitonic fivebrane solution of $D=10$ supergravity \cite{Lu}
appears as a magnetic monopole \cite{Hmono} or a string \cite{Khuri2} according
as it wraps around $5$ or $4$ of the compactified directions \footnote{It could
in principle also appear as a membrane by wrapping around $3$ of the
compactified
directions, but the $N=4$ supergravity theory (\ref{3})
obtained by naive dimensional reduction does not admit the membrane
solution \cite{Khuri2}.}.
Regarding this dual string as fundamental in its own right interchanges the
roles of $T$-duality and $S$-duality. The solitonic monopole states obtained
in this way thus play the same role for the dual string as the elementary
electric winding states play for the fundamental string. The Kaluza-Klein
states are common to both. Since these solitons are extreme ($a= \sqrt 3$)
black holes \cite{Khuri}, however, it follows by $S$-duality that the
elementary Kaluza-Klein states should be black holes too! By $T$-duality,
the same holds true of the elementary winding states. Rather than invoke
$S$-duality, however, we shall proceed directly to establish that the
elementary states described above are in one-to-one correspondence with the
extreme electric black holes\footnote{The idea that there may be a dual theory
which interchanges Kaluza-Klein states and Kaluza-Klein monopoles was
previously discussed in the context of $N=8$ supergravity by Gibbons and
Perry \cite{Perry}}. Now this leaves open the possibility that they  have
the same masses and quantum numbers but different interactions. Although we
regard this possibility as unlikely given the restrictions of $N=4$
supersymmetry, the indirect argument may be more compelling in this respect
(even though it suffers from the drawback that $S$-duality has not yet been
rigorously established).  Of course, elementary states are supposed to be
singular and solitonic states non-singular. How then can we interchange
their roles? The way the theory accommodates this requirement is that when
expressed in terms of the fundamental metric $e^{a\phi}g_{\mu\nu}$ that
couples to the worldline of the superparticle the elementary solutions are
singular and the solitonic solutions are non-singular, but when expressed in
terms of the dual metric $e^{-a\phi}g_{\mu\nu}$, it is the other way around
\cite{Lu3,Khuri}.

Let us begin by recalling the bosonic sector of the four dimensional action
for the massless fields obtained by dimensional reduction from the usual
($2$-form) version of $D=10$ supergravity:
\bea
\lefteqn{ S=\frac{1}{32\pi}\int d^4x\sqrt{-G}e^{-\Phi}[R_G +
G^{\mu\nu}\partial_{\mu}\Phi\partial_{\nu}\Phi
-\frac{1}{12}G^{\mu\lambda}
 G^{\nu\tau}G^{\rho\sigma}H_{\mu\nu\rho}H_{\lambda\tau\sigma}}\nn\\&&
-\frac{1}{4}G^{\mu\lambda}G^{\nu\tau}F_{\mu\nu}{}^a(LML)_{ab}
   F_{\lambda\tau}{}^b+ \frac{1}{8}
     G^{\mu\nu}Tr(\partial_{\mu}ML\partial_{\nu}ML)]
\la{3}
\eea
where $F_{\mu\nu}{}^a=\partial_{\mu}A_{\nu}{}^a-\partial_{\nu}A_{\mu}{}^a$
and
$H_{\mu\nu\rho}=(\partial_{\mu}B_{\nu\rho}+2A_{\mu}{}^aL_{ab}F_{\nu\rho}{}^b)
+ {\rm permutations}$.  Here $\Phi$ is the $D=4$ dilaton, $R_G$ is the
scalar curvature formed from the string metric $G_{\mu\nu}$, related to the
canonical metric $g_{\mu\nu}$ by $G_{\mu\nu}\equiv e^{\Phi}g_{\mu\nu}$.
$B_{\mu\nu}$ is the 2-form which couples to the string worldsheet and
$A_{\mu}{}^a$ ($a=1,...,28$) are the abelian gauge fields. $M$ is a
symmetric $28\times28$ dimensional matrix of scalar fields satisfying
$MLM=L$ where $L$ is the invariant metric on $O(6,22)$:
\be
L=\pmatrix{0&I_6&0\cr I_6&0&0 \cr 0&0&-I_{16}}.
\la{L}
\ee
The action is invariant under the $O(6,22)$ transformations
$M\rightarrow\Omega M\Omega^T$,
$A_{\mu}{}^a\rightarrow\Omega^{a}{}_{b}A_{\mu}{}^b$, $G_{\mu\nu}\rightarrow
G_{\mu\nu}$, $B_{\mu\nu}\rightarrow B_{\mu\nu}$, $\Phi\rightarrow\Phi$,
where $\Omega$ is an $O(6,22)$ matrix satisfying $\Omega^TL\Omega=L$.
$T$-duality corresponds to the $O(6,22;Z)$ subgroup and is known to be an
exact symmetry of the full string theory. The equations of motion, though
not the action, are also invariant under the $SL(2,R)$ transformations:
${\cal M}\rightarrow \omega{\cal M}\omega^T,{\cal F}_{\mu\nu}{}^{a\alpha}
  \rightarrow  \omega^{\alpha}{}_{\beta}{\cal F}_{\mu\nu}{}^{a\beta},
 g_{\mu\nu}\rightarrow g_{\mu\nu},\,\,M\rightarrow M$
where $\a=1,2$ with ${\cal F}_{\mu\nu}{}^{a1}=F_{\mu\nu}{}^{a}$ and ${\cal
F}_{\mu\nu}{}^{a2}=\left(\lambda_2(ML)^a{}_{b}\tilde F_{\mu\nu}{}^{b}+
\lambda_1 F_{\mu\nu}{}^{a}\right)$, where $\omega$ is an $SL(2,R)$ matrix
satisfying  $\omega^T{\cal L}\omega={\cal L}$ and where
\be
{\cal M}=\frac{1}{\lambda_2}\left(\begin{array}{cc}
1&\lambda_1\\
\lambda_1&|\lambda|^2
\end{array}\right),\,\,\,
{\cal L}=\left(\begin{array}{cc}
0&1\\
-1&0\end{array}\right).
\la{4}
\ee
$\l$ is given by $\lambda=\Psi+ie^{-\Phi}\equiv\lambda_1+i\lambda_2$. The
axion $\Psi$ is defined through the relation
$\sqrt{-g}H^{\mu\nu\rho}=-e^{2\Phi}
\epsilon^{\mu\nu\rho\sigma}\partial_{\sigma}\Psi$.  $S$-duality corresponds
to the $SL(2,Z)$ subgroup and there is now a good deal of evidence
\cite{Schwarz} in favor of its also being an exact symmetry of the full
string theory. For the restricted class of configurations obtained by
setting to zero the $16$ gauge fields $F^{13\rightarrow28}$ originating from
the ten-dimensional gauge fields, it is possible to define a dual action
\cite{Schwarz} which has manifest $SL(2,R)$ symmetry. The field strengths
$F^{1\rightarrow6}$, whose origin resides in the $D=10$ metric, remain the
same but the  $F^{7\rightarrow12}$, whose origin resides in the $D=10$
$2$-form, are replaced by their duals. The equations of motion are also
invariant under $O(6,6)$; the action is not except for the $SL(6,R)$
subgroup which acts trivially. This action is precisely the one obtained by
dimensional reduction from the dual ($6$-form) version of $D=10$
supergravity which couples to the worldvolume of the fivebrane \cite{Duff1}
and for which the axion is just the $6$-form component lying in the extra
$6$ dimensions.  This provides another reason for believing that the roles
of $S$ and $T$ duality are interchanged in going from string to fivebrane
\cite{Schwarz,Binetruy,Khuri2}
and is entirely consistent with an earlier observation that the dual theory
interchanges the worldsheet and spacetime loop expansions \cite{Lu2}. In
this light, the need to treat the above $16$ gauge fields on a different
footing is only to be expected since in the dual formulation their kinetic
terms are $1$-loop effects \cite{Lu2}.

We now turn to the electric and magnetic charge spectrum. Schwarz
and Sen \cite{Schwarz} present an $O(6,22;Z)$ and $SL(2,Z)$ invariant
expression for the mass of particles saturating the strong Bogomol'nyi bound
$m=|Z_1|=|Z_2|$:
\be
m^2=\frac{1}{16}(\alpha^a~~\beta^a){\cal M}^{0}(M^{0}+L)_{ab}
    \left(\begin{array}{c}
\alpha^b\\
\beta^b\end{array}\right)
\la{5}
\ee
where a superscript $0$ denotes the constant asymptotic values of the
fields. Here $\alpha^a$ and $\beta^a$ ($a=1,...,28$) each belong to an even
self-dual Lorentzian lattice $\Lambda$ with metric given by $L$ and are
related to the electric and magnetic charge vectors $(Q^a,P^a)$ by
$(Q^a,P^a)=\left(M_{ab}{}^0(\alpha^b + \lambda_{1}{}^0\beta^b)/\lambda_2{}^0,
L_{ab}\beta^b\right)$.  As discussed in \cite{Schwarz} only a subset of the
conjectured spectrum corresponds to elementary string states.  First of all
these states will be only electrically charged, i.e. $\b =0$, but there
will be restrictions on $\a$ too. Without loss of generality let us focus on
a compactification with $M{}^0=I$ and $\lambda_{2}{}^{0}=1$. Any other toroidal
compactifications can be brought into this form by $O(6,22)$ transformations
and a constant shift of the dilaton.  The mass formula (\ref{5}) now becomes
\be
m^2=\frac{1}{16 }
  {\alpha}^a (I+L)_{ab}  {\alpha}^b=
 \frac{1}{8}
 \left({\alpha}_R \right)^2
\la{7}
\ee
with $ { \alpha}_R = \frac{1}{2} (I+L) { \alpha}$ and
${ \alpha}_L = \frac{1}{2} (I-L) { \alpha}$. In the string
language ${\a}_{R(L)}$ are the right(left)-moving internal momenta.
The mass of a generic string state in the Neveu-Schwarz sector (which is
degenerate with the Ramond sector) is given by
\be
m^2=\frac{1}{8\lambda_2{}^{0}} \left\{
 \left({ \alpha}_R \right)^2 +2 N_R -1 \right\}=
 \frac{1}{8 \lambda_2{}^{0}} \left\{
\left({ \alpha}_L \right)^2 +2 N_L -2 \right\}.
\label{8}
\ee
A comparison of (\ref{7}) and
(\ref{8}) shows that the string states
satisfying the Bogomol'nyi bound all have $N_R=1/2$. One then finds
\be
 N_L-1=\frac{1}{2} \left(\left( { \alpha}_R \right)^2
 -\left( { \alpha}_L \right)^2\right)
 =\frac{1}{2} { \alpha}^{T} L { \alpha},
\label{9}
\ee
leading to ${ \alpha}^{T} L { \alpha}\geq -2$.  We shall now show that
extreme black holes with $a=\sqrt{3}$ are string states with ${
\alpha}^{T} L { \alpha}$ null ($N_L=1$) and those with $a=1$ are string
states with ${ \alpha}^{T} L { \alpha}$ spacelike $(N_L>1)$. We have
been unable to identify solutions of the low-energy field equations (\ref{3})
corresponding to states with ${ \alpha}^{T} L { \alpha}$ timelike
$(N_L<1)$. \footnote{In the {\it non-abelian} theory Sen \cite{Schwarz}
identifies these states with the electric analogues of BPS monopoles.}

 Let us first focus on the $a=\sqrt{3}$ black hole. To identify it as a
state in the spectrum we have to find the corresponding charge vector ${
\a}$ and to verify that the masses calculated by the formulas (\ref{2}) and
(\ref{5}) are identical. The action (\ref{3}) can be consistently truncated
by keeping the metric $g_{\mu\nu}$, just one field strength ($F=F^1$, say),
and one scalar field $\phi$ via the ansatz $\Phi=\phi/\sqrt{3}$ and
$M_{11}=e^{2\phi/\sqrt{3}}=M_{77}^{-1}$. All other diagonal components of
$M$ are set equal to unity and all non-diagonal components to zero. Now
(\ref{3}) reduces to (\ref{1}) with $a=\sqrt{3}$. (This yields the electric
and magnetic Kaluza-Klein (or "F") monopoles. This is not quite the
truncation chosen in \cite{Khuri}, where just $F^7$ was retained and
$M_{11}=e^{-2\phi/\sqrt{3}}=M_{77}^{-1}$. This yields the the electric and
magnetic winding (or "H") monopoles.  However, the two are related by
$T$-duality).  We shall restrict ourselves to the purely electrically
charged solution with charge $Q=1$, since this one is expected to correspond
to an elementary string excitation. The charge vector ${ \a}$ for this
solution is obviously given by $\alpha^a=\delta^{a,1}$ with ${
\alpha}^{T} L { \alpha}=0$. Applying (\ref{5}) for the mass of the state
we find $m^2=1/16=Q^2/16$, which coincides with (\ref{2}) in the extreme
limit.  This agreement confirms the claim that this extreme $a=\sqrt{3}$
black hole is a state in the Sen-Schwarz spectrum and preserves $2$
supersymmetries.

Next we turn to the $a=1$ black hole. The theory is consistently truncated
by keeping the metric, $F=F^1=F^7$ and setting $M=I$. The only non-vanishing
scalar is the dilaton $\Phi\equiv\phi$.  Now (\ref{3}) reduces to (\ref{1})
with $a=1$ but $Q^2=2$. An extreme $a=1$ black hole with electric charge Q
is then represented by the charge configuration
$\alpha^a=\delta^{a,1}+\delta^{a,7}$. Applying (\ref{5}) we find
$m^2=1/4=Q^2/8$ which coincides with (\ref{2}) in the extreme limit.
Therefore the $a=1$ extreme solution is also in the spectrum, and has ${
\a}^{T} L { \a}=2$ or $N_L=2$.

Although physically very different, we can see with hindsight that both the
$a=\sqrt3$ and $a=1$ black holes permit a uniform mathematical treatment by
noting that both may be obtained from the Schwarzschild solution by
performing an $[O(6,1) \times O(22,1)]/[O(6) \times O(22)$ transformation
\cite{Sen2}. The $28$ parameters of this transformation correspond to the
$28$ $U(1)$ charges.  If $\gamma$ and ${ u}$ correspond to the boost
angle  and a 22 dimensional unit vector respectively, associated with
O(22,1)/O(22)  transformations, $\delta$ and ${ v}$ denote the boost
angle and the 6  dimensional unit vector respectively, associated with the
O(6,1)/O(6) transformations, and $m_0$ is the mass of the original
Schwarzschild black hole, then the mass and charges of the new black hole
solution are given by \cite{Sen2}:
\[m ={1\over 2} m_0 (1+\cosh\gamma \cosh\delta)\]
\[\alpha_L ={\sqrt 2} m_0 \cosh\delta \sinh\gamma \, { u}\]
\be
 \alpha_R ={\sqrt 2} m_0 \cosh\gamma \sinh\delta \, { v}
\la{9a}
\ee
(Note that the convention about $R$ and $L$ of \cite{Sen2} is opposite to
the one used in the present paper). Black holes with $\alpha^T L\alpha=0$
are generated by setting $\gamma=\delta$, whereas black holes with $\alpha^T
L\alpha>0$ are generated by setting $\gamma<\delta$. The Bogomol'nyi bound
given in (\ref{7})  corresponds to $m^2=( \alpha_R)^2/8$. This bound is
saturated by taking the limit where the mass $m_0$ of the original
Schwarzschild black hole approaches 0 and the parameter $\delta$ approaches
$\infty$, keeping the product $m_0\sinh\delta$ fixed. As  discussed in
\cite{Sen2}, this is precisely the extremal limit. Thus we see that extremal
black holes satisfy the Bogomol'nyi relation, both for $\alpha^T L \alpha=0$
and $\alpha^T L\alpha>0$.

{}From the above $a=\sqrt 3$ solution we can generate the whole set of
supersymmetric black hole solutions with ${ \alpha}^{T} L { \alpha}=0$
in the following way: first we note that we are interested in constructing
black hole solutions with different charges but with fixed asymptotic values
of $M$ (which here has been set to the identity).  Thus we are not allowed
to make $O(6,22)$ transformations that change the asymptotic value of $M$.
This leaves us with only an $O(6) \times O(22)$ group of transformations.
The effect of these transformations acting on the parameters given in
(\ref{9a}) above is to transform the vectors ${ u}$ and ${ v}$ by
$O(22)$ and $O(6)$ transformations respectively without changing the
parameters $\gamma$ and $\delta$.  Now, the original $a=\sqrt 3$ solution
corresponds to a choice of parameters $\gamma=\delta$, ${
u}^m=\delta_{m1}$ and ${ v}^m =\delta_{k1}$. It is clear that an $O(6)
\times O(22)$ transformation can rotate ${ u}$ and ${ v}$ to arbitrary
$22$ and $6$ dimensional unit vectors respectively, without changing
$\gamma$ and $\delta$. Since this corresponds to the most general charge
vector satisfying $\alpha^T L\alpha=0$, we see that the $O(6) \times O(22)$
transformation can indeed generate an arbitrary black hole solution with
$\alpha^TL\alpha=0$ starting from the original $a=\sqrt 3$ solution. This
clearly leaves the mass invariant, but the new charge vector ${ \a}'$
will in general not be located on the lattice. To find a state in the
allowed charge spectrum we have to rescale ${ \a}'$ by a constant $k$ so
that ${ \a}''=k{ \a}'$ is a lattice vector. Clearly the masses
calculated by (\ref{2}) and (\ref{5}) still agree (this is obvious by
reversing the steps of rotation and rescaling), leading to the conclusion
that all states obtained in this way preserve 1/2 of the supersymmetries.
Therefore all states in the spectrum belonging to $s_{max}=1$
supermultiplets for which $N_R=1/2,N_L=1$ are extreme $a=\sqrt{3}$ black
holes.

Let us now turn to the case of the $a=1$ solution. In this case the original
solution corresponds to the choice of parameters $\gamma=0$, ${
v}^m=\delta_{m1}$. (For $\gamma=0$, the parameter ${ u}$ is irrelevant).
An $O(6) \times O(22)$ transformation can rotate ${ v}$ to any other 6
dimensional unit vector, but it cannot change the parameters $\delta$ and
$\gamma$. As a result, the final solution will continue to have $\gamma=0$
and hence $\alpha_L=0$.  Since this does not represent the most general
charge vector $\alpha$, with $\alpha^T L \alpha>0$, we see that the most
general black hole representing states with $\alpha^T L\alpha>0$ is not
obtained in this way even after rescaling. The missing  states with
$\a_L\neq 0$  are constructed by choosing $\gamma$  so that $\tanh ^2
\gamma={\a_L^2}/{\a_R^2}$, and ${ u}, \ { v}$ as for the $a=\sqrt{3}$
case, followed by a suitable $O(6) \times O(22)$ rotation. Clearly, those
solutions  are extreme black holes too. However, for these solutions  a
truncation to an effective action of the form (\ref{1}) is not possible.
The following picture arises: for a fixed value of $\a_R^2$, $\a_L^2$ can
vary in the range $\a_R^2\geq \a_L^2\geq 0$. The boundary states are
described by the well-known $a=\sqrt{3}$ ($\a_R^2=\a_L^2$) and $a=1$
($\a_L^2=0$) black holes, whereas the states in between  cannot be related
to a single scalar-Maxwell parameter $a$. But all solutions preserve 1/2 of
the supersymmetries.

 It should also be clear that the purely magnetic extreme black hole
solutions \cite{Khuri} obtained from the above by the replacements $
\phi\rightarrow -\phi, \alpha\rightarrow\beta$ will also belong to the
Schwarz-Sen spectrum of solitonic states.  Starting from either the purely
electric or purely magnetic solutions, dyonic states in the spectrum which
involve non-vanishing axion field $\Psi$ can then be obtained by $SL(2,Z)$
transformations. Specifically, a black hole with charge vector $({
\alpha}, 0)$ will be mapped into ones with charges $(a{ \alpha}, c{
\alpha})$ with the integers $a$ and $c$ relatively prime \cite{Schwarz}.

Not all black hole solutions of (\ref{3}) belong to the Sen-Schwarz
spectrum, however. Let us first consider the Reissner-Nordstrom solution.
Since this black hole solves the equations of $N=2$ supergravity, whose
bosonic sector is pure Einstein-Maxwell, it solves (\ref{3}) as well.  The
required consistent truncation is obtained by keeping $g_{\mu\nu}$,
$F=F^1=F^7=\tilde F^2=\tilde F^8$ and setting $\Phi=0$, $M=1$. Now (\ref{3})
effectively reduces to (\ref{1}) with $a=0$ but $Q^2=4$. On the other hand,
if it were in the Schwarz-Sen spectrum its charge vectors would be given by
$\alpha^a=\delta^{a,1}+\delta^{a,7}$ with ${ \alpha}^{T}L { \alpha}=2$
and  $\beta^a=\delta^{a,2}+\delta^{a,8}$ with ${ \beta}^{T}L{
\beta}=2$.  Applying (\ref{5}) for the mass of the state we find $m^2=1/2$,
which disagrees with the result $m^2=1$ obtained from the extreme limit of
(\ref{2}). So the test fails and the $a=0$ black hole does not belong to the
Schwarz-Sen spectrum.  This was only to be expected since it breaks $3/4$ of
the supersymmetries and hence saturates the weaker Bogomol'nyi bound
$m=|Z_1|,|Z_2| =0$ \cite{Kallosh}.  Such black holes belong to the $32$
complex dimensional ($s_{max}=3/2$) supermultiplet. We see no reason to
exclude these states from the full string spectrum, however. Another example
of a black hole solution not in the Schwarz-Sen spectrum is the $a=1$
dilaton black hole of \cite{Garfinkle} where the only non-vanishing gauge
field is $F^{13}$. This has mass $m^2=Q^2/8$ but according to (\ref{5}) its
mass would vanish.  Again, this contradiction is only to be expected since
this solution breaks all the supersymmetries, in contrast with the
$F=F^1=F^7$ embedding discussed above. We do not know whether such black
holes saturating no Bogomol'nyi bound ($m>|Z_1|,|Z_2|)$, which include the
neutral Schwarzschild black holes ($Z_1=Z_2=0$), are also in the string
spectrum. States with these quantum numbers would belong to the 256
dimensional ($s_{max}\geq 2$) supermultiplets. According to \cite{Gibbons2},
however, black holes breaking all the supersymmetries do not themselves form
supermultiplets. This would appear to contradict the claim that {\it all}
black holes are string states.

In the supersymmetric case, all values of $a$ lead to extreme black holes
with zero entropy but their temperature is zero, finite or infinite
according as $a<1$, $a=1$ or $a>1$, and so in \cite{Holzhey} the question
was posed: can only $a>1$ scalar black holes describe elementary particles?
We have not definitively answered this question but a tentative response
would be as follows.  First we note that the masses and charge vectors are
such that the lightest $a=0$ black hole may be regarded as a bound state
(with zero binding energy) of two lightest $a=1$ black holes which in turn
can each be regarded as bound states (again with zero binding energy) of two
lightest $a=\sqrt{3}$ black holes. Thus if by elementary particle one means
an object which cannot be regarded as a bound state, then indeed extreme
scalar black holes with $a>1$ are the only possibility, but if one merely
means a state in the string spectrum then $a\leq1$ extreme scalar black
holes are also permitted.

We have limited ourselves to $N_R=1/2$ supermultiplets with $s_{min}=0$.
Having established that the $s=0$ member of the multiplet is an extreme
black hole, one may then use the fermionic zero modes to perform
supersymmetry transformations to generate the whole supermultiplet of black
holes \cite{Gibbons2,Aichelburg} with the same mass and charges. Of course,
there are $N_R=1/2$ multiplets with $s_{min}>0$ coming from oscillators with
higher spin and our arguments have nothing to say about whether these are
also extreme black holes. They could be naked singularities. Indeed,
although in this paper we have focussed primarily on identifying certain
massive heterotic string states with extreme black holes, perhaps equally
remarkable is that these elementary string states can be described at all by
solutions of the supergravity theory.  In a {\it field} theory, as opposed
to a {\it string} theory, one is used to having as elementary massive states
only the Kaluza-Klein modes with $s_{max}=2$. However, as we have already
seen, the  winding states (usually thought of as intrinsically stringy) are
on the same footing as Kaluza-Klein states as far as solutions are
concerned, so perhaps the same is true for the $s>2$ states.

None of the spinning $N_R=1/2$ states is described by extreme {\it rotating}
black hole metrics because they obey the same Bogomol'nyi bound as the
$s_{min}=0$ states, whereas the mass formula for an extreme {\it rotating}
black hole depends on the angular momentum $J$. Moreover, it is the fermion
fields which carry the spin in the $s_{min}=0$ supermultiplet. (For the
$a=0$ black hole, they yield a gyromagnetic ratio $g=2$ \cite{Aichelburg};
the $a=\sqrt{3}$ and $a=1$ superpartner $g$-factors are unknown to us).  It
may be that there are states in the string spectrum described by the extreme
rotating black hole metrics but if so they will belong to the $N_R\neq 1/2$
sector\footnote{The gyromagnetic and gyroelectric ratios of the states in
the heterotic string spectrum would then have to agree with those of charged
rotating black hole solutions of the heterotic string. This is indeed the
case: the $N_L=1$ states \cite{Hosoya} and the rotating $a=\sqrt{3}$ black
holes \cite{Wiltshire} both have $g=1$ whereas the $N_L>1$ states
\cite{Russo} and the rotating $a=1$ \cite{Sen} (and $a=0$ \cite{Schild})
black holes both have $g=2$. In fact, it was the observation that the Regge
formula $J\sim m^2$ also describes the mass/angular momentum relation of an
extreme rotating black hole which first led Salam \cite{Salam} to imagine
that elementary particles might behave like black holes!}.  Since, whether
rotating or not, the black hole solutions are still independent of the
azimuthal angle and independent of time, the supergravity theory is
effectively {\it two-dimensional} and therefore possibly integrable. This
suggests that the spectrum should be invariant under the larger duality
$O(8,24;Z)$ \cite{Duff}, which combines $S$ and $T$. The corresponding
Kac-Moody extension would then play the role of the spectrum generating
symmetry \cite{Geroch}.

Conversations with G. Gibbons, R. Khuri and J. Liu are gratefully acknowledged.
\pagebreak

\end{document}